# Some Generalizations of the Capacity Theorem for AWGN Channels

K.K. Sharma

*Abstract*—The channel capacity theorem for additive white Gaussian noise channel (AWGN), widely known as the Shannon-Hartley Law, expresses the information capacity of a channel bandlimited in the conventional Fourier domain in terms of the signal-to-noise ratio in it. In this letter generalized versions of the Shannon-Hartley Law using the linear canonical transform (LCT) are presented. The channel capacity for AWGN channels is found to be a function of the LCT parameters.

I. INTRODUCTION

The channel information capacity formula, widely known as the Shannon-Hartley law [1]-[2], expresses the information capacity of a power-limited additive white Gaussian noise (AWGN) channels bandlimited in the conventional Fourier domain (CFD) as a function of the signal-to-noise ratio in it. The assumption of the strictly bandlimited channel implies that the associated impulse response is non-zero almost everywhere on $(-\infty, \infty)$ and this raises the realizability and other issues [2] for achieving the channel capacity. It is also known that the capacity is achievable only when the time of transmission $T$ tends to infinity [2]. The channel capacity calculations for a more realistic scenario employing signals belonging to a wider class that are approximately time-limited and/or approximately bandlimited in the CFD has also been done in [2]. The channel capacity of the root-mean square (RMS) bandlimited gaussian channels using time-limited signals (such as PAM signals) has also been derived in [3].

Recently much interest has been shown in the linear canonical transform (LCT) [5], [6], [11], [12], and it has been shown that the LCT outperform the conventional Fourier transform (CFT) in many applications [4]-[7], [11]. The issue of compact support and bandlimitedness of signals in the LCT and CFT domains has been discussed in [4], and [7]. Specifically, it has been shown in [4], [7] that a signal bandlimited in the CFD can also be bandlimited in the LCT domains with different bandwidths in each domain.

Similarly, it has been shown in [11] that the sampling rate for perfect reconstruction for the class of signals bandlimited in the LCT domains can be much lower than the Nyquist rate. Moreover, it is quite possible for the class of signals considered in [2] and [3] for the purpose of the transmission over the channel (that are not strictly bandlimited in the CFD) to be bandlimited in the LCT domains.

These findings immediately motivates one to think whether these results will have implications on the channel capacity for bandlimited channels which revolves around the issue of bandlimitedness of signals and the sampling theory of bandlimited signals [1], [2]. Moreover, the following questions remain unanswered in the existing theory of bandlimited AWGN channels. Can a signal having infinite bandwidth in the CFD be transmitted over a strictly bandwidth AWGN channel? Can a strictly bandlimited signal in the CFD be transformed to some other signal so that it can be transmitted over an AWGN channel with smaller bandwidth. We answer these questions in this letter using the theory of LCT and propose some generalizations of the Shannon-Hartley law involving the channel capacity calculations under AWGN condition using signals bandlimited in the CFD and LCT domains.

The rest of the paper is organized as follows. In section II we provide a brief review of the LCT and the fractional Fourier transform. In section III we present the main results of the letter in the generalizations of the Shannon-Hartley law for signals bandlimited in the CFT and LCT domains. The conventional Shannon-Hartley law can be seen as a special case of the result presented here. The paper is concluded in Section IV.

## II. REVIEW OF THE LCT AND FRFT

The fractional Fourier transform (FRFT) and the LCT are generalizations of the CFT [2] and have been shown to be more useful than the CFT in many signal-processing applications [5]-[10]. The LCT of a signal $f(t)$ with parameter matrix $M = [A, B; C, D]$, denoted as $F_M(u)$, is given by [5]

$$F_M(u) \triangleq [\Re_M f](u) = \begin{cases} \sqrt{\dfrac{1}{j2\pi B}} \int_{-\infty}^{\infty} f(t) K_M(u,t) dt & \text{for } B \neq 0 \\ \sqrt{D} \exp(jCDu^2/2) f(Du) & \text{for } B = 0 \end{cases}, \quad (1)$$

where $K_M(t,u) = \exp\left[\frac{j}{2}\left(\frac{D}{B}u^2 - 2\frac{1}{B}ut + \frac{A}{B}t^2\right)\right]$, $B \neq 0$, and the determinant of the matrix $M$ satisfies the relation $AD - BC = 1$. Various signal transforms such as the Fresnel transform and the FRFT are simply special cases of the LCT. The FRFT is in fact a one-parameter subgroup of the group of LCTs. To be specific, if we substitute $A = D = \cos a$, and $B = -C = \sin a$ in the matrix $M$ and evaluate (1), we obtain the FRFT of the signal within a multiplicative constant. Moreover, the FRFT reduces to the CFT for $a = p/2$ [5].

The signal $f(t)$ can also be expressed in terms of the signal $F_M(u)$ using the inverse LCT relation given by [5]

$$f(t) = \begin{cases} \int_{-\infty}^{\infty} \sqrt{\frac{1}{-j2pB}} F_M(u) K_M^*(t,u) du, & B \neq 0 \\ \sqrt{A} \exp(-jCAt^2/2) F_M(At), & B = 0 \end{cases}. \quad (2)$$

The other properties and comprehensive discussion of the LCT can be seen in [2], [3], and the references therein.

### III. GENERALIZED INFORMATION CAPACITY THEOREMS FOR AWGN CHANNELS

In this section we present the main results of the paper as a generalizations of the celebrated Shannon-Hartley law [1] using the LCT.

#### A. GENERALIZED CHANNEL CAPACITY THEOREM FOR BANDLIMITED SIGNALS IN CFD:

***Theorem 1:*** The information capacity of the modified Shannon's time-continuous AWGN channel bandlimited to $W$ Hertz in the CFD as shown in Fig. 1 with LCT parameter matrix $\tilde{M} = [B^{-1}, -D; 0, B]$, $B \neq 0$, is given by

$$I_{C,W} = \frac{W}{B} \log_2\left(1 + \frac{P}{hW}\right) \text{ bits/second.} \quad (3)$$

where $I_{C,W}$, $P$, and $h/2$ denote the information capacity of the channel, average transmitted power, and the power spectral density (PSD) of the AWGN respectively.

***Proof:*** The celebrated Shannon-Hartley law defines the fundamental limit on the information capacity of error-free transmission of a power limited Gaussian continuous channel transmitting a zero-mean stationary process $X(t)$ bandlimited to $W$ Hertz in the CFD and it can be expressed as [1], [2]:

$$I_C = W \log_2\left(1 + \frac{P}{hW}\right) \text{ bits/second,} \quad (4)$$

where $I_C, P$, and $\mathbf{h}/2$ denote the information capacity of the channel, average transmitted power and the power spectral density (PSD) of the AWGN respectively. Also the information capacity per transmission is given by [1]

$$I_C = \frac{1}{2}\log_2\left(1 + \frac{P}{hW}\right) \text{bits/transmission,} \qquad (5)$$

The eqn. (4) is obtained from (5) by multiplying it with the Nyquist rate of sampling for bandlimited signals in the CFD, i.e., $2W$ transmissions per second. The rate in (4) can be achieved by the transmission of the sinc functions ($\sin(\mathbf{p}t)/\mathbf{p}t$) at a rate of $2W$ transmissions per second as discussed in [2] but this requires infinite delay for achieving it because of the infinite support of the sinc functions in time domain. But it is well known that a signal $f(t)$ bandlimited to $W$ in the CFD will also have compact support $W_M = WB$ in the LCT domain with parameter matrix $M = [A, B; C, D]$ for the specific value of the parameter $A = 0$, i.e., $F_M(u)$ has compact support $W_M = WB$ [4]. Therefore we propose the modified Shannon' time-continuous model shown in Fig.1 and let the signal given to the channel be expressed as:

$$s_i(t) = \sum_{m=1}^{N} x_{im} \frac{\sin(2\mathbf{p}W(t - m/2W))}{2\mathbf{p}W(t - m/2W)}, \qquad (6)$$

where $\sum_{m=1}^{N} x_{im}^2 \leq 2WPT$, $i = 1, 2, ..., K$ and $N = 2WT$. Here $K = 2^{R_bT}$ denotes the total number of messages and $R_b$ is the rate at which the message source emits binary digits/second and messages are presented to the channel after every $T$ seconds and

$$X_i(t) = \sum_{m=1}^{N} x_{im}\mathbf{d}(t - m/2W).$$

On taking the LCT with parameter matrix $\tilde{M}$ of (5), we obtain

$$\left[\Re_{\tilde{M}} s_i\right](t) = \sum_{m=1}^{N} x_{im} \Re_{\tilde{M}}\left(\frac{\sin(2\mathbf{p}W(t - m/2W))}{2\mathbf{p}W(t - m/2W)}\right)(t), \qquad (7)$$

where $M = M_{\mathbf{p}/2}\tilde{M}$ and $M = [0, B; -B^{-1}, D], B \neq 0$.

The conventional bandwidth of the signal $\Re_{\tilde{M}}\left(\frac{\sin(2\mathbf{p}Wt)}{2\mathbf{p}Wt}\right)(t)$ and hence of the signal $\left[\Re_{\tilde{M}} s_i\right](t)$ can be easily shown to be $WB$ [4]. The signal given in (6) can be obtained from (7) by taking the LCT with parameter matrix $\tilde{M}^{-1}$. The rest of the procedure for recovering $x_{im}$, $m = 1, 2, ..., N$ from (7) is identical to that described in [2] for AWGN channel bandlimited in the CFD. In effect we have transmitted $N$ real numbers in time $T$. Thus, along the lines of [2], the generalized channel capacity theorem for the AWGN channel having a bandwidth $WB$ shown in Fig. 1 an can be expressed as

$$I_{C,WB} = W \log_2\left(1 + \frac{P}{hWB}\right) \text{ bits/second}. \tag{8}$$

Therefore, the capacity of the channel having a bandwidth $W$ can be written as given in (3). ∎

For this channel, the generalized Shannon limit of $\frac{E_b}{h}$ for $W = \infty$, i.e., $\left(\frac{E_b}{h}\right)_\infty$ can be easily shown to be $\left(\frac{E_b}{h}\right)_\infty = \frac{\ln 2}{B}$. It is clear from (8) that the channel capacity tends towards infinity as $B$ tends towards zero.

To derive another version of the channel capacity theorem, we again consider the modified Shannon' time-continuous model shown in Fig.1.

***Theorem 2:*** The information capacity of the modified Shannon's time-continuous AWGN channel bandlimited to $W$ Hertz in the CFD as shown in Fig. 1 with LCT parameter matrix $M = \left[A, B; 0, A^{-1}\right], A \neq 0$, is given by

$$I_{C,W} = AW \log_2\left(1 + \frac{P}{hW}\right) \text{ bits/second}. \tag{9}$$

***Proof:*** The signal given to the channel is expressed as in (6). The LCT with parameter matrix $M = \left[A, B; 0, A^{-1}\right], A \neq 0$ of (5) is given by

$$[\Re_M s_i](t) = \sum_{m=1}^{N} x_{im} \Re_M\left(\frac{\sin(2pW(t - m/2W))}{2pW(t - m/2W)}\right)(t), \tag{10}$$

The conventional bandwidth of the signal $\Re_M\left(\frac{\sin(2pWt)}{2pWt}\right)(t)$ can be shown to be $W/A$ [4]. The signal given in (6) can be obtained from (10) by taking the LCT with parameter matrix $M^{-1} = \left[A^{-1}, -B; 0, A\right]$. The rest of the procedure for recovering $x_{im}$, $m = 1, 2, ..., N$ from (10) is identical to that described in [2]. Thus the generalized channel capacity for an AWGN channel shown in Fig. 1 having a bandwidth $W/A$ can be expressed as

$$I_{C,W/A} = W \log_2\left(1 + \frac{AP}{hW}\right) \text{ bits/second.} \tag{11}$$

Therefore, the generalized channel capacity theorem for the channel having a bandwidth $W$ can be expressed as given in (9). ∎

It is clear from (8) and (11) that the channel capacity can be increased above the usual value of the capacity given in (3) by proper selection of the LCT parameters in it. The generalized Shannon limit of $\frac{E_b}{h}$ for $W = \infty$, i.e., $\left(\frac{E_b}{h}\right)_\infty$ for the (11) can be shown to be $\left(\frac{E_b}{h}\right)_\infty = A \ln 2$.

### B. CHANNEL RATE TO TRANSMIT SIGNALS BANDLIMITED IN THE LCT DOMAINS:

***Theorem 3:*** The information capacity of the modified Shannon's time-continuous AWGN channel bandlimited to $W$ Hertz in the CFD as shown in Fig. 2 with LCT parameter matrix $M = [A, B; C, D], A \neq 0$, is given by

$$I_{C,W} = W \log_2\left(1 + \frac{P(4WT - 1)}{hW}\right) \text{ bits/second} \tag{12}$$

***Proof:*** Consider again the modified Shannon' time-continuous model shown in Fig.2 and let the signal given to the channel be expressed as

$$s_i(t) = \sum_{m=1}^{N} x_{im} f(t - m/2W), \tag{13}$$

where the signal $f(t)$ is the impulse response of the system whose input is the signal $X_i(t) = \sum_{m=1}^{N} x_{im} d(t - m/2W)$. The signal $f(t)$ is assumed to have compact support $W_M$ in some LCT domain $M = [A, B; C, D], A \neq 0$. The LCT with parameter matrix $\tilde{M}$ of (13) is then given by

$$\left[\Re_{\tilde{M}} s_i\right](u) = \sum_{m=1}^{N} x_{im} \Re_{\tilde{M}}\left(f(t - m/2W)\right)(u), \tag{14}$$

where $M = M_{p/2}\tilde{M}$. Now using the delay property given in [5] and keeping in mind the compact support of the signal $f(t)$ in the LCT domain $M = [A, B; C, D], A \neq 0$, it can be shown that the conventional spectrum of the signal $\left[\Re_{\tilde{M}} s_i\right](u)$ consists of $N = 2WT$ copies of $\left[\Re_M f(t)\right](u)$ and the minimum bandwidth of the signal without any overlap between these copies is given by $W = (2N-1)W_M$ provided the parameter of the LCT satisfy $A \geq 4WW_M$. The sequence $x_{im}$, $m = 1, 2, ..., N$ can be obtained by filtering each copy of $[\Re_M f(t)](u)$ from the spectrum of the signal $[\Re_{\tilde{M}} s_i](u)$. The channel capacity, keeping in mind the bandwidth of each copy of $[\Re_M f(t)](u)$ (which happens to be $2W_M$), can then be expressed as

$$I_{C,W} = W \log_2 \left(1 + \frac{P}{\mathbf{h}W_M}\right) \text{ bits/second.} \tag{15}$$

The eqn. (15) can also be expressed as given in (12). ∎

Once again the channel capacity tends towards infinity as $T$ tends to infinity. It may be reiterated, however, that channel models proposed here also suffers from all the problems mentioned in [2] for the Shannon's classical model.

**C. TRANSMISSION OF A SIGNAL BANDLIMITED IN THE CFD IN A CHANNEL HAVING ARBITRARY BANDWIDTH:**

Having derived some generalized versions of the Shannon Hartley Law, in this subsection we attempt to answer the questions raised in the introduction of this paper, i.e., we compute the channel capacity required for the transmission of a signal bandlimited in the CFD. We again state the result that a signal $f(t)$ bandlimited to $W$ in the CFD will also have compact support $W_M = WB$ in the LCT domain with parameter matrix $M = [A, B; C, D]$ for the specific value of the parameter $A = 0$, i.e., $F_M(u)$ has compact support $W_M = WB$ [4]. A bandlimited signal in CFD $f(t)$ can be transmitted by transmitting its samples in time domain at a rate of $2W$ samples /second or by the samples of the LCT of the signal $f(t)$ in the domain with parameter matrix $\tilde{M}$ at a different rate of $2W_M$ samples /second. Here the LCT domain with parameter matrix $\tilde{M}$ is related to (i.e., samples of the signal $F_{\tilde{M}}(u)$) the LCT domain $M$ as given by

$$M = M_{p/2} \tilde{M}, \tag{16}$$

where $M_{p/2} = [0, 1; -1, 0]$ is the matrix corresponding to the CFT. The required sampling rate in this case is given by the well-known Nyquist rate $u_s \geq 2W_M = 2WB$. Thus the information capacity of the channel of bandwidth $W_M$ from (3) is given by

$$I_{C,W_M} = WB \log_2 \left(1 + \frac{P}{\mathbf{h}WB}\right) \text{ bits/second.} \tag{17}$$

The expression (17) for the capacity can be maximized with respect to the parameter $B$ of the LCT. The maximum value of the capacity can be obtained by solving the equation

$$\ln(1 + (P/\mathbf{h}WB)) = P/(P + \mathbf{h}WB), \tag{18}$$

for the optimum value of the parameter $B_{opt}$ and substituting it in (17) to yield the result

$$\left(I_{C,W_M}\right)_{max} = \max_B I_{C,W_M} = 1.44 W B_{opt} \frac{P}{P + hWB_{opt}}. \tag{19}$$

The equation (19) can also be written in terms of the channel bandwidth $W_{M_{opt}}$ as

$$\left(I_{C,W_M}\right)_{max} = 1.44 W_{M_{opt}} \frac{P}{P + hW_{M_{opt}}}. \tag{20}$$

It is obvious from (20) that for a given bandwidth of the channel, the capacity can be different from the value given by (3). Moreover, the bandwidth of the channel for its transmission can be either larger or smaller than the bandwidth $W$ of the signal $f(t)$ in the CFD. The Eqn. (20) for the case of $W_{M_{opt}} = \infty$ reduces to the well-known formula for the channel capacity $\left(I_C\right)_\infty$ for infinite bandwidth [1]:

$$\left(I_C\right)_\infty = 1.44 \frac{P}{h} \text{ bits/second}. \tag{21}$$

**D. TRANSMISSION OF A SIGNAL WITH INFINITE BANDWIDTH IN CFD BUT BANDLIMITED IN THE LCT DOMAINS:**

Now we compute the channel capacity required for the transmission of the signals having infinite bandwidth in the CFT domain but whose FRFT with angle $a$ has compact support in some LCT domain [11], [4]. For such signals, it has been shown in [8] that the sampling rate $u_s$ for perfect signal reconstruction for the class of signals bandlimited in the LCT domains to $W_M$ is given by

$$u_s \geq \frac{2W_M}{(A \sin a + B \cos a)}. \tag{22}$$

The information transmitted by such samples is given by

$$I_{C,W_M} = \frac{W_M}{(A \sin a + B \cos a)} \log_2 \left(1 + \frac{P(A \sin a + B \cos a)}{hW_M}\right) \text{ bits/second}. \tag{23}$$

It is clear that (3) is a special case of (23) for $a = 0$ and $B = \sin p/2$, i.e., for signals bandlimited in the CFD. However, it may be mentioned here that a signal can be bandlimited in more than one LCT domains simultaneously as opposed to the case of a signal bandlimited in FRFT domains [4], [7]. Thus for a given signal, the channel information capacity depends on the LCT parameters chosen for the purpose of signal transmission. The expression (23) for the capacity can be maximized with respect to the parameters $A, B, a$ of the LCT and FRFT. The optimum value set of the parameters can be obtained by solving the equation

$$\ln\left(1+(P/hWB'_{opt})\right) = P/(P+hWB'_{opt}), \tag{24}$$

where $B'_{opt} = (A\sin a + B\cos a)$. And the maximum value of the capacity can be shown to be

$$\left(I_{C,W_M}\right)_{max} = \max_{A,B,a} I_{C,W_M} = 1.44 \frac{W}{B'_{opt}}\left(\frac{P}{P+\left[hW/B'_{opt}\right]}\right) \text{bits/second.} \tag{25}$$

It may be noted that for the value of $W = \infty$, (25) also reduces to (21).

### III. CONCLUSIONS

The generalized Shannon-Hartley Law or the generalized information capacity theorems for bandlimited channels using the LCT theory are presented. The information capacity is seen to be functions of the FRFT and/or LCT parameters. It is also shown that it is possible to transmit some class of signals having infinite bandwidth in CFD but bandlimited in the LCT domains in a channel with finite capacity.

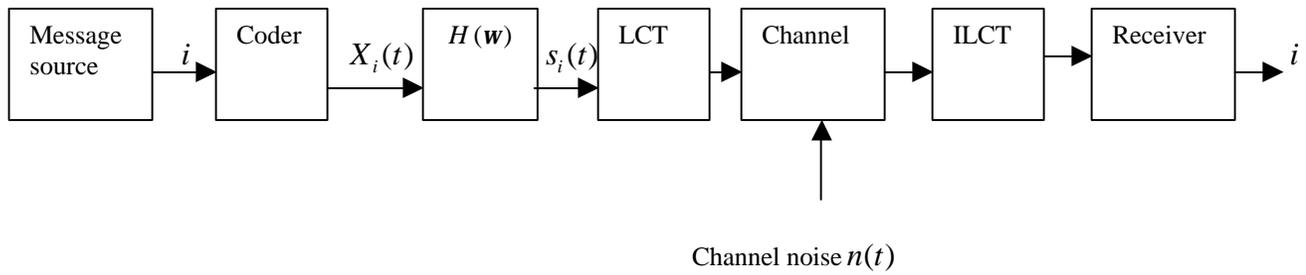

Fig. 1: The modified Shannon's time-continuous channel model: The coder output is LCTed and additive noise is added to it. The output of the channel is lowpass filtered followed by the inverse LCT operation.

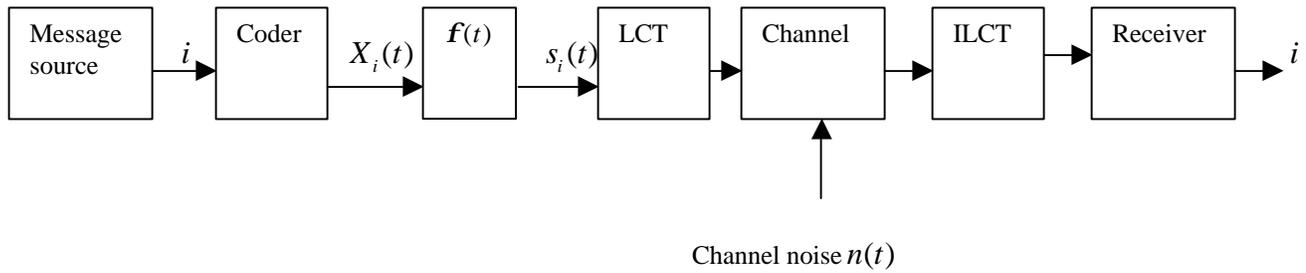

Fig. 2: The modified Shannon's time-continuous channel model: The coder output is LCTed and additive noise is added to it. The output of the channel is lowpass filtered followed by the inverse LCT operation.